\documentclass[12pt]{article}
\usepackage{graphicx,color}
\usepackage{XoohmE}
\usepackage{booktabs}
\def\bo{{\raise.005ex\hbox{\large$\Box$}}}

 \parskip=\medskipamount                 
 \thicklines                         

 \def\bpl{\Big(}
 \def\bpr{\Big)}
 
 \def\ve{\varepsilon}
 
 \def\der{\partial}
 \def\brr{\begin{equation}}
 \def\err{\end{equation}}
 \def\brr{\begin{eqnarray}}
 \def\err{\end{eqnarray}}
 \def\ba{\left(\begin{array}}
 \def\ea{\end{array}\right)}
 \def\lf{\left.\begin{array}{c}}
 \def\rf{\end{array}\right.}

 \newcommand{\dr}{\raise.3ex\hbox{$\stackrel{\leftarrow}{\partial }$}{}}
 \newcommand{\dl}{\raise.3ex\hbox{$\stackrel{\rightarrow}{\partial}$}{}}
 \newcommand{\topi}{\raise.3ex\hbox{$\stackrel{\pi}{\longrightarrow}$}{}}

 \def\qd{{\kern0.5pt
                   q \kern-5.05pt \raise5.8pt\hbox{$\textstyle.$}\kern 0.5pt}}

\def\4{\oplus}
\def\8{\otimes}

\def\spin{\mathfrak{spin}}

\definecolor{Hey}{rgb}{.9,.05,.4}
\definecolor{orange}{rgb}{1,.5,0}
\definecolor{plum}{rgb}{.4,0,.6}
\definecolor{R}{rgb}{1,0,0}
\definecolor{G}{rgb}{0,1,0}
\definecolor{B}{rgb}{0,0,1}
\font\cfnt=lcircle10 at 9pt
\def\lplus{\mathop{\kern2pt
            \raise1.275ex\hbox to0pt{\cfnt\char"07\hss}\kern-.6pt+}}
\def\YT#1#2{\vcenter{\hbox{\vbox{\baselineskip0pt\parskip=\medskipamount%
             \def\B{$\sqcap$\llap{$\sqcup$}\kern-1.9pt}
              \def\Bd{\hbox{\kern2.4pt\raise.4pt\hbox{$\cdot$}\kern-5.7pt\B\kern0pt}}
              \def\4{\raise.25pt\hbox to0pt{\hss\kern2pt--\hss}}
              \def\Z{\hfil\vskip-5.9pt}\lineskiplimit0pt\lineskip0pt%
               \setbox0=\hbox{#1}\hsize\wd0\parindent=0pt#2}\,}}}
\newdimen\parshift\parshift=\parindent
\catcode`@=11
 \long\def\@footnotetext#1{\insert\footins{\reset@font\footnotesize\interlinepenalty%
  \interfootnotelinepenalty\splittopskip\footnotesep\splitmaxdepth\dp\strutbox%
   \floatingpenalty\@MM\hsize\columnwidth\addtolength{\hsize}{-2\parindent}
    \@parboxrestore\protected@edef\@currentlabel{\csname p@footnote\endcsname\@thefnmark}
      \color@begingroup
       \@makefntext{\rule\z@\footnotesep\ignorespaces#1\@finalstrut\strutbox}
        \color@endgroup}}
 \long\def\@makefntext#1{\hglue\parshift
                         \vbox{\noindent\hb@xt@0em{\hss\@makefnmark\,}#1}}
\catcode`@=12
 \def\bpl{\Big(}
 \def\bpr{\Big)}
 \def\ba{\left(\begin{array}}
 \def\ea{\end{array}\right)}
 
 \def\der{\partial}
 \def\brr{\begin{eqnarray}}
 \def\err{\end{eqnarray}}

 \def\S{\mathbb S}

 \newcommand{\fr}[2]{{\textstyle\frac{#1}{#2}}}

 \HarvTitles 
 \seceq

\def\be{\begin{equation}}
\def\ee{\end{equation}}
\def\bea{\begin{eqnarray}}
\def\eea{\end{eqnarray}}

 \def\qd{{\kern0.5pt
                   q \kern-5.05pt \raise5.8pt\hbox{$\textstyle.$}\kern 0.5pt}}

\begin{document}

 {\hbox to\hsize{July 2009 \hfill {SUNY-O/702}}}

 \vspace{.5in}

 \begin{center}
 {\Large\bf Spin Holography via Dimensional Enhancement}
 \\[.25in]
 {Michael G. Faux$^a$ and Gregory D. Landweber$^b$}\\[3mm]
 {\small\it
  $^a$ Department of Physics,
      State University of New York, Oneonta, NY 13820\\[-1mm]
  {\tt  fauxmg@oneonta.edu}
  \\
 $^b$Department of Mathematics, Bard College, Annandale-on-Hudson, NY
 12504-5000\\[-1mm]
  {\tt gregland@bard.edu}
 }\\[9mm]
 {\bf ABSTRACT}\\[.01in]
 \begin{quotation}
 {\it We explain how all information about ambient component field spin assignments
 in higher-dimensional off-shell supersymmetry is
 accessibly coded in one-dimensional restrictions, known as shadows.  We also explain
 how to determine whether the components of a given one-dimensional supermultiplet
 may assemble into representations of $\spin(1,D-1)$ and, if so, how to specifically
 determine those representations.}
 \end{quotation}
 \end{center}

 In a recent paper \cite{Enhancement} we discussed how the representation
 theory of supersymmetry in diverse dimensions lies encoded within the
 representation theory of one-dimensional supersymmetry.  This
 hinges on the recognition that each supersymmetric
 field theory has a unique analog, called its shadow, obtained by
 restricting to a zero-brane.  In our earlier paper, we exhibited
 an algebraic ``litmus test" which determines whether or not a given
 one-dimensional theory represents a shadow of another
 theory in a larger specified number of ``ambient" dimensions.
 We call the (re)construction of a higher-dimensional field theory based on information
 encoded as a one-dimensional shadow mechanics ``dimensional enhancement".
 This describes the reverse of dimensional reduction. In our earlier
 work we used our new technology to reproduce known results about
 the representations of four-dimensional $N=1$ supersymmetry using
 reasoning exclusively contextualized to one-dimension.  In this
 letter we extend this process, explaining how to
 resolve the organization of multiplet component fields into
 specific representations of $\spin(1,D-1)$ based only on
 one-dimensional muliplet structures.

 Our approach to this problem is enabled by casting the algebraic
 requirements of supersymmetry in terms of certain ``linkage
 matrices", or their graphical equivalents known as Adinkra diagrams
 \cite{FG1}.
 This approach is complementary to traditional methods based on
 Salam-Strathdee superfields, and enables new forms of analysis.
 The linkage matrices are
 generalizations of the ``garden matrices" introduced in
 \cite{GatesRana}, while the feasibility, realized in \cite{Enhancement}, that generic
 supersymmetries might be universally encoded by one-dimensional superalgebras
 was conjectured in \cite{Enuf}.\footnote{Other novel approaches to
 this problem have also been developed, \eg, as explained in \cite{T2}.}

 The physical motivation which enabled us to derive our enhancement
 condition was the observation, described in \cite{Enhancement},
 that the linkgage matrices describing any
 supersymmetric field theory must be invariant under spin transformations.
 This is analogous to the statement that Dirac Gamma matrices should be
 Lorentz invariant, in the context of generic (not necessarily
 supersymmetric) field theories with
 spinors. Indeed,
 it is easy to show that Gamma matrices are invariant objects;
 the proof relies on the structure of the relevant Clifford algebras.
 Our method generalizes this observation to the case of
 supersymmetric theories.  In this case Clifford algebras also play a
 central role.  The reason for this is that the linkage matrices satisfy
 conditions, equivalent to the algebraic statements of supersymmetry,
 which generalize the ``garden algebras" introduced
 in \cite{GatesRana}, which are themselves variants of Clifford algebras.

 In \cite{Enhancement} we presented a proof that spin invariance
 correlates all space-like linkage matrices in terms of time-like linkage
 matrices, and we also showed how spin invariance implies
 constraints on the $\spin(1,D-1)$-representations spanned by the
 multiplet component fields. In that paper we focussed on the
 implications of the former observation, and we did not fully
 resolve the implications concerning the specific organization
 of component fields into spin representations.  In the bulk of this
 letter we do precisely that.\footnote{A more mathematically-precise discussion
 complementary to much of our analysis is provided in
 \cite{Filtered}.}

 Supersymmetry is generated by first-order linear differential operators
 which are also matrices acting on the vector spaces spanned by the
 component bosons $\phi_i$ and the component fermions
 $\psi_{\hat{\imath}}$.  Specifically, the supercharges are written as
 \brr (\,Q_A\,)_i\,^{\hat{\imath}} &=&
      (\,u_A\,)_i\,^{\hat{\imath}}
      +(\,\Delta^\mu_A\,)_i\,^{\hat{\imath}}\,\der_\mu
      \nonumber\\[.1in]
      (\,\tilde{Q}_A\,)_{\hat{\imath}}\,^i &=&
      i\,(\,\tilde{u}_A\,)_{\hat{\imath}}\,^i
      +i\,(\,\tilde{\Delta}^\mu_A\,)_{\hat{\imath}}\,^i\,\der_\mu
      \,,
 \label{ssm}\err
 where $u_A$, $\tilde{u}_A$, $\Delta^\mu_A$, and
 $\tilde{\Delta}^\mu_A$ are real valued ``linkage matrices", and
 the index $A$ is an ambient spinor index.  We consider here the
 important case where the linkage matrices are ``Adinkraic", which
 means that each of these has only one non-vanishing entry in each
 row and one non-vanishing entry in each column.  Moreover, each
 non-vanishing entry assumes one of the two values $\pm 1$. As we
 explain in \cite{Enhancement}, these cases correspond to all known
 supermultiplets in dimensions $D>1$, and is arguably non-restrictive.
 An explanation of how these matrix structures may be generically
 achieved by linear restructuring of the components
 is explained mathematically in \cite{DFGHIL01}.

 The component fields assemble into representations of
 $\spin(1,D-1)$, whereby the generating transformations
 act as
 $\delta\,\phi_i=\fr12\,\theta^{\mu\nu}\,(\,T_{\mu\nu}\,)_i\,^j\,\phi_j$
 and $\delta\,\psi_{\hat{\imath}}=
 \fr12\,\theta^{\mu\nu}\,(\,\tilde{T}_{\mu\nu}\,)_{\hat{\imath}}\,^{\hat{\jmath}}\,
 \psi_{\hat{\jmath}}$, where $\theta^{0a}$ parameterizes a boost in the $a$-th spatial
 direction and $\theta^{ab}$ parameterizes a rotation in the
 $ab$-plane. Our goal is to explain how these
 representations are determined from the linkage matrices described
 above.

 In Appendix A of \cite{Enhancement} we presented a simple
 derivation that $\spin(1,D-1)$-invariance of the bosonic linkage
 matrices $\Delta^\mu_A$ and $u_a$ translates into the following six
 constraints,
 \brr \Delta^a_A &=&
      -\fr12\,(\,\Gamma^0\Gamma^a\,)_A\,^B\,
      \Delta^0_B
       -T^{0a}\,\Delta^0_A
      +\Delta^0_A\,\tilde{T}^{0a}
      \nonumber\\[.1in]
      \fr12\,(\,\Gamma_{ab}\,)_A\,^B\,\Delta^0_B
      &=&
      \Delta^0_A\,\tilde{T}_{ab}
      -T_{ab}\,\Delta^0_A
      \nonumber\\[.1in]
      \delta_b\,^a\,\Delta^0_A &=&
      \fr12\,(\,\Gamma_0\Gamma_b\,)_A\,^B\,\Delta^a_B
      +T_{0b}\,\Delta^a_A
      -\Delta^a_A\,\tilde{T}_{0b}
      \nonumber\\[.1in]
      \eta^{a[b}\,\Delta^{c]}_A
      +\fr14\,(\,\Gamma^{bc}\,)_A\,^B\,\Delta^a_B
      &=& \fr12\,\Delta^a_A\,\tilde{T}^{bc}
      -\fr12\,T^{bc}\,\Delta^a_A
      \nonumber\\[.1in]
      \fr12\,(\,\Gamma_0\Gamma_a\,)_A\,^B\,u_B &=&
      u_A\,\tilde{T}_{0a}
      -T_{0a}\,u_A
      \nonumber\\[.1in]
      \fr12\,(\,\Gamma_{ab}\,)_A\,^B\,u_B &=&
      u_A\,\tilde{T}_{ab}
      -T_{ab}\,u_A
       \,.
 \label{t1}\err
 These are a direct translation of equations (A.4), (A.5), and
 (A.7) from that paper, where the component index structures such as
 $(\cdot)_i\,^{\hat{\imath}}$ have been suppressed, and where the
 boost and rotation operators have been separated.
 The requirement that the fermion linkage matrices
 $\tilde{\Delta}^\mu_A$ and $\tilde{u}_A$ be
 $\spin(1,D-1)$-invariant imposes a set of constraints similar to
 (\ref{t1}).  These ``fermionic analogs" are obtained from
 (\ref{t1}) by adding tildes to the linkage matrices and by
 exchanging the boson and fermion spin generators,
 via $T_{\mu\nu}\leftrightarrow \tilde{T}_{\mu\nu}$.
 There is significant redundancy in the equations (\ref{t1}) and the
 corresponding fermionic analogs.
 One of our
 goals in this letter is to distill these statements to their
 essence.

 In this letter we focus on multiplets in a ``standard" configuration,
 according to which $\Delta^0_A=\tilde{u}_A^T$ and $\tilde{\Delta}^0_A=u_A^T$.
 All non-gauge matter multiplets may be construed, using judicious linear
 re-organization of components, so that the linkage matrices have this
 feature. In terms of Adinkras, these
 identities say that every Adinkra edge codifies both an ``upward directed" supersymmetry
 transformation and a corresponding ``downward directed" supersymmetry transformation.
 The presence of gauge
 structures modifies the situation in a way which is tractable, but which adds distracting technical
 details. In the interest of keeping things as simple as possible, but maintaining
 generality for the wide class of interesting multiplets described by non-gauge
 matter, we utilize this feature.  We are developing the general case, which includes
 gauge structures, in on-going work. We use also the features
 that boost generators are described by symmetric matrices, $T_{0a}=T_{0a}^T$,
 and that rotation generators are described by anti-symmetric matrices, $T_{ab}=-T_{ab}^T$.
 \footnote{The symmetry properties of the spin generators were addressed in
 footnote 20 in \cite{Enhancement}, and can be proved using standard
 representation theory.}

 We now address the matter of removing redundancies from the list of
 constraints given in (\ref{t1}).  First,
 the matrix transposes of the fifth equation in (\ref{t1})
 and its fermionic analog tell us
 \brr \fr12\,(\,\Gamma_0\Gamma_a\,)_A\,^B\,\tilde{\Delta}^0_A &=&
      \tilde{T}_{0a}\,\tilde{\Delta}^0_A
      -\tilde{\Delta}^0_A\,T_{0a}
      \nonumber\\[.1in]
      \fr12\,(\,\Gamma_0\Gamma_a\,)_A\,^B\,\Delta^0_A &=&
      T_{0a}\,\Delta^0_A
      -\Delta^0_A\,\tilde{T}_{0a} \,,
 \label{tran}\err
 where we have used the features $\Delta^0_A=\tilde{u}_A^T$ and $\tilde{\Delta}^0_A=u_A^T$
 described above.  (These identities allow us to cast the entirety of the
 spin invariance requirements in terms of the ``down" matrices
 $\Delta^\mu_A$ and $\tilde{\Delta}^\mu_A$.)

 It proves helpful that
 the boost generators $T_{0a}$ and $\tilde{T}_{0a}$ appear in (\ref{tran}) in
 the same combination as in the first equation in (\ref{t1}).
 Accordingly, we can use (\ref{tran}) to eliminate these terms,
 to derive a constraint independent of the component spin
 representation assignments.  Specifically, if we
 substitute (\ref{tran})  into
 the first equations of (\ref{t1}) and its fermionic analog, we
 easily determine
 \brr \Delta^a_A &=&
      -(\,\Gamma^0\Gamma^a\,)_A\,^B\,\Delta_B^0
      \nonumber\\[.1in]
      \tilde{\Delta}^a_A &=&
      -(\,\Gamma^0\Gamma^a\,)_A\,^B\,\tilde{\Delta}_B^0 \,.
 \label{g1}\err
 This was discussed in \cite{Enhancement} where the equation
 (\ref{g1}) enabled the first important ``sieve" in the enhancement
 problem.  This is an interesting result because this demonstrates
 that the space-like linkage matrices are completely determined by
 the time-like linkage matrices, and also because this result is
 disconnected from the spin representation assignments of the component
 fields.

 It is easy to verify that only the first, second, and fifth equation
 in (\ref{t1}) describe independent conditions, and that the third,
 fourth, and sixth equations are satisfied automatically given
 these, along with the ``standard" conditions $\tilde{\Delta}^0_A=u_A^T$
 and $\Delta^0_A=\tilde{u}_A^T$.
 Specifically, the transpose of the fermionic analog of the sixth equation in (\ref{t1})
 is identical to the second equation in (\ref{t1}).
 Next, if we substitute equations (\ref{tran}) and (\ref{g1}) into the
 third equation in (\ref{t1}), we find this is satisfied
 automatically.
 Finally, if we substitute the second equation in
 (\ref{t1}) and equation (\ref{g1}) into the fourth
 equation in (\ref{t1}), we find that this is satisfied automatically.
 We have already replaced the first equation in (\ref{t1}) with
 (\ref{g1}).  This leaves only the second and the fifth equations in (\ref{t1})
 as independent conditions, the latter of which has already been replaced with
 (\ref{tran}).

 For purposes of being helpfully specific, we restrict our focus to a
 four-dimensional ambient space.\footnote{Cases involving
 diverse dimensions can be dealt with similarly.  However,
 the general case requires notational wizardry which we think
 detracts from the elegance of our result, at least for this introductory
 presentation.}
 We also streamline our notation as follows.  We designate $B_a:=T_{0a}$ and $\tilde{B}_a:=\tilde{T}_{0a}$ as
 the generators of boosts in the $a$-th spatial direction as
 realized on
 the bosons and fermions, respectively.\footnote{The boost operator $B_a$ should not be confused with the magnetic
 field components assigned a similar name in \cite{Enhancement}.}
 And we designate $R^a:=\fr12\,\ve^{abc}\,T_{bc}$ and
 $\tilde{R}^a=\fr12\,\ve^{abc}\,\tilde{T}_{bc}$ as the generators of rotations about
 the $a$-th coordinate axis as realized on the bosons
 and fermions, respectively.  Moreover, the boost generator acting on ambient spinors
 is ${\cal B}_a=\fr12\,\Gamma_0\Gamma_a$,
 and the rotation generator acting on ambient spinors is
 ${\cal R}_a=\fr14\,\ve_{abc}\,\Gamma^{bc}$.
 Finally, we write the time-like boson and fermion ``down" linkage matrices
 respectively as $d_A=\Delta^0_A$ and
 $\tilde{d}_A=\tilde{\Delta}^0_A$, whereby the ``standard" symmetry properties
 become $u_A=\tilde{d}_A^T$ and $\tilde{u}_A=d_A^T$.

 Using the notational refinements described in the previous paragraph,
 we can rewrite (\ref{tran}) and its fermionic analog as
 \brr (\,{\cal B}_a\,)_A\,^B\,(\,d_B\,)_i\,^{\hat{\imath}}
      &=&
      (\,d_A\,)_i\,^{\hat{\jmath}}\,(\,\tilde{B}_a\,)_{\hat{\jmath}}\,^{\hat{\imath}}
      -(\,B_a\,)_i\,^j\,(\,d_A\,)_j\,^{\hat{\imath}}
      \nonumber\\[.1in]
      (\,{\cal B}_a\,)_A\,^B\,(\,u_B\,)_i\,^{\hat{\imath}}
      &=&
      -(\,u_A\,)_i\,^{\hat{\jmath}}\,(\,\tilde{B}_a\,)_{\hat{\jmath}}\,^{\hat{\imath}}
      +(\,B_a\,)_i\,^j\,(\,u_A\,)_j\,^{\hat{\imath}} \,.
 \label{b1}\err
 Note that the second equation in (\ref{b1}) may be obtained from the first by toggling the
 placement of tildes, taking a matrix transpose, using the
 standard relationship $\tilde{d}_A^T=u_A^T$, and the property that
 the boost generators are symmetric matrices.

 Similarly, we can rewrite the second equation in (\ref{t1}) and its fermionic analog
 as
 \brr  (\,{\cal R}_a\,)_A\,^B\,(\,d_B\,)_i\,^{\hat{\imath}}
      &=&
      (\,d_A\,)_i\,^{\hat{\jmath}}\,(\,\tilde{R}_a\,)_{\hat{\jmath}}\,^{\hat{\imath}}
      -(\,R_a\,)_i\,^j\,(\,d_A\,)_j\,^{\hat{\imath}}
      \nonumber\\[.1in]
      (\,{\cal R}_a\,)_A\,^B\,(\,u_B\,)_i\,^{\hat{\imath}}
      &=&
      (\,u_A\,)_i\,^{\hat{\jmath}}\,(\,\tilde{R}_a\,)_{\hat{\jmath}}\,^{\hat{\imath}}
      -(\,R_a\,)_i\,^j\,(\,u_A\,)_j\,^{\hat{\imath}}
      \,.
 \label{r1}\err
 Note that the second equation in (\ref{r1}) may be obtained from the first by toggling the
 placement of tildes, taking a matrix transpose, using the
 standard relationship $\tilde{d}_A^T=u_A^T$, and the property that
 the rotation generators are anti-symmetric matrices.

 Note that equations (\ref{b1}) and (\ref{r1}) describe the full
 implication implied by the Lorentz invariance of the linkage matrices
 aside from the requirement (\ref{g1}).  Since (\ref{g1}) does not
 involve the component spin representation matrices, it follows that
 (\ref{b1}) and (\ref{r1}) represent the distillation we had sought.

 For a given Adinkra the matrices $d_A$ and $u_A$ are
 readily determined.  Moreover, as explained in \cite{Enhancement}, we lose no
 generality by selecting an ambient spin basis, whereby
 the matrices $(\,{\cal B}_a\,)_A\,^B$ and $(\,{\cal R}_a\,)_A\,^B$ are also
 determined.  This reduces (\ref{b1}) and (\ref{r1})
 into a linear algebra problem cast as a system of matrix equations
 in terms of the 12 yet-undetermined matrices
 $B_a$, $R_a$, $\tilde{B}_a$, and $\tilde{R}_a$.\footnote{This counting
 is specific to the case of four-dimensions, of course.}
 The linear algebra problem posed by (\ref{b1}) and (\ref{r1}) is
 tractable in all cases.  By this we mean that if solutions exist,
 these may be readily obtained via routine systematic algorithmic methods.
 But in certain important cases, this problem actually admits a
 closed-form solution.  It is interesting to explain one such
 circumstance in detail, before commenting on the general case.

 Consider the case where the bosons are Lorentz scalars, whereby $B_a=0$ and
 $R_a=0$.\footnote{The four-dimensional $N=1$ Chiral multiplet
 is a characteristic example.  Other examples with more supersymmetries, corresponding
 to analogs of hypermultiplets without central charges, were identified in \cite{RETM}.}
 Under this circumstance, we can
 derive a closed-form solution for the component spin generators
 $\tilde{B}_a$ and $\tilde{R}_a$, which act on the fermion fields.
 To do this we reorganize (\ref{b1}) and (\ref{r1})
 into equivalent versions obtained by taking sums and differences of the
 first and second equations in each case. In this way we derive
 \brr (\,{\cal B}_a\,)_A\,^B\,(\,d_B\pm u_B\,)_i\,^{\hat{\imath}} &=&
      (\,d_A\mp u_A\,)_i\,^{\hat{\jmath}}\,(\,\tilde{B}_a\,)_{\hat{\jmath}}\,^{\hat{\imath}}
      \nonumber\\[.1in]
      (\,{\cal R}_a\,)_A\,^B\,(\,d_B\pm u_B\,)_i\,^{\hat{\imath}} &=&
      (\,d_A\pm
      u_A\,)_i\,^{\hat{\jmath}}\,(\,\tilde{R}_a\,)_{\hat{\jmath}}\,^{\hat{\imath}}
      \,.
 \label{stan}\err
 By considering separately the upper and lower choices for the
 signs, these describe four groups of equations, each group
 with three equations --- one for each choice of the spatial
 coordinate index $a$ --- for each of the $N$ choices of the spinor
 index $A$.\footnote{Note that $N$ counts the number of
 one-dimensional supersymmetries.}
 Thus, these describe $2\times 3\times 4\times\,N=24\,N$ matrix
 equations.

 \begin{table}
 \begin{center}
 \begin{tabular}{cc}
 $u_1 \,\,=\,\, \ba{cc|cc}1&&&\\&1&&\\\hline &&0&\\&&&0\ea$ &
 $d_1 \,\,=\,\, \ba{cc|cc}0&&&\\&0&&\\\hline &&1& \\&&&1 \ea$
 \\[.5in]
 $u_2 \,\,=\,\, \ba{cc|cc}&1&&\\-1&&&\\\hline &&&0\\&&0& \ea$ &
 $d_2 \,\,=\,\, \ba{cc|cc} &0&&\\0&&&\\\hline &&&-1\\&&1& \ea$
 \\[.5in]
 $u_3 \,\,=\,\, \ba{cc|cc} &&-1& \\&&&-1\\\hline 0&&&\\&0&& \ea$ &
 $d_3 \,\,=\,\, \ba{cc|cc} &&0& \\&&&0\\\hline 1&&&\\&1&& \ea$
 \\[.5in]
 $u_4 \,\,=\,\, \ba{cc|cc}&&&-1\\&&1& \\\hline &0&&\\0&&& \ea$ &
 $d_4 \,\,=\,\, \ba{cc|cc} &&&0\\&&0& \\\hline &-1&&\\1&&& \ea$
 \end{tabular}
 \caption{Linkage matrices corresponding to the shadow of the four-dimensional $N=1$ Chiral multiplet.}
 \label{links242}
 \end{center}
 \end{table}

 As it turns out, for standard Adinkras, the matrices
 $d_A\pm u_A$ are in all cases non-singular.\footnote{These combinations also
 describe garden matrices \cite{GatesRana,Enuf}.}
 Moreover, these combinations describe Clifford actions, and each such
 combination describes a matrix which squares to plus or minus the
 identity.  These assertions are easy to prove using the
 algebra defined by (2.7) and (2.8) in \cite{Enhancement}.

 \pagebreak

 \noindent
 The fact that $d_A\pm u_A$ are non-singular allows us to
 ``solve" (\ref{stan}) by writing
 \brr (\,\tilde{B}_a\,)_{\hat{\imath}}\,^{\hat{\jmath}} &=&
      \fr{1}{N}\,(\,{\cal B}_a\,)_{A}\,^B\,\bpl\,
      (\,d^A\pm u^A\,)^{-1}\,(\,d_B\mp
      u_B\,)\,\bpr_{\hat{\imath}}\,^{\hat{\jmath}}
      \nonumber\\[.1in]
      (\,\tilde{R}_a\,)_{\hat{\imath}}\,^{\hat{\jmath}} &=&
      \fr{1}{N}\,(\,{\cal R}_a\,)_A\,^B\,\bpl\,
      (\,d^A\pm u^A\,)^{-1}\,(\,d_B\pm
      u_B\,)\,\bpr_{\hat{\imath}}\,^{\hat{\jmath}} \,.
 \label{mz}\err
 To obtain this result we matrix multiply both sides of
 (\ref{stan})
 from the left by the matrix inverse of $d_A\pm u_A$ for any
 particular choice of the index $A$.  Since we must obtain the same
 result for $(\,\tilde{B}_a\,)_{\hat{\imath}}\,^{\hat{\jmath}}$ and
 $(\,\tilde{R}_a\,)_{\hat{\imath}}\,^{\hat{\jmath}}$ for each separate choice of
 the index $A$, it follows that
 we can add a sum over $A$ and then divide by $N$, since the index
 $A$ takes $N$ values.  But this implies several
 checks on the ability to consistently define the matrices
 $\tilde{B}_a$ and $\tilde{R}_a$.  In particular, each of the $N$ terms
 in the $A$-sums in (\ref{mz}) needs to describe the same matrix
 for each choice of the ambiguous sign and for each choice of the index $a$.  In total, this imposes
 $3\,(\,2\,N-1\,)$ conditions for each of the two equations in (\ref{mz}), for
 a total of $6\,(\,2\,N-1\,)$ extra ``spin-consistency"
 conditions.\footnote{This counting is specific to a four-dimensional ambient space.}

 In \cite{Enhancement} we derived consistency conditions
 describing a ``first sieve" which distinguishes whether
 postulate enhancements of one-dimensional supermultiplets
 properly close the higher-dimensional supersymmetry algebra.
 In that paper we also described a ``second sieve", corresponding to the
 spin-statistics theorem, according to which fermions must assemble as
 spinors and bosons as tensors.  The consistency conditions described
 above would seemingly impose a third sieve, corresponding to the
 requirement that the component fields resolve into
 representations of $\spin(1,D-1)$.

 As a helpful example, we consider the particular case of the shadow of the four-dimensional
 ${\cal N}=4$ Chiral multiplet. In \cite{Enhancement} we showed that this corresponds to a
 particular 2-4-2 Adinkra equivalent to the specific linkage matrices shown in Table
 \ref{links242}.  Presently, we use these linkage matrices as a
 starting point, and proceed to use the above formalism to resolve
 the spin representation assignments of the fermions,
 using the assumption that the bosons enhance to scalars.  We will
 show that our formalism properly predicts how the fermion vertices
 assemble as an ambient spacetime spinor.  In this case we know that
 the one-dimensional multiplet enhances because we obtained this
 also by dimensional reduction, as explained in Appendix C of
 \cite{Enhancement}.  But our analysis here is blind to this fact,
 so that our analysis provides a built-in consistency check on the
 very formalism that we are developing.  With this in mind, it is
 gratifying that the following analysis properly accounts for the
 four-dimensional spin structure.

 We use the particular 4D spin structure implied by the
 Majorana basis described in Appendix A of
 \cite{Enhancement}.  In this basis, the ambient spinor boost and rotation
 generators ${\cal B}_a$ and ${\cal R}_a$ are exhibited in
 Table \ref{boorots}.
 Using these matrices, and using the linkage matrices in Table \ref{links242}, we can apply (\ref{mz})
 to address our spin-consistency condition, \ie, the third seive, and
 then use (\ref{mz}) to actually compute the component spin structure.
 For example, using the matrix $(\,{\cal B}_1\,)_A\,^B$ specified in Table \ref{boorots},
 the first equation in (\ref{mz}) implies
 \brr \tilde{B}_1 &=& \fr18\,(\,d^1-u^1\,)^{-1}\,(\,d_3+u_3\,)
      +\fr18\,(\,d^3-u^3\,)^{-1}\,(\,d_1+u_1\,)
      \nonumber\\[.1in]
      & &
      +\fr18\,(d^2-u^2\,)^{-1}\,(\,d_4+u_4\,)
      +\fr18\,(d^4-u^4\,)^{-1}\,(\,d_2+u_2\,) \,.
 \label{b1test}\err
 It is easy to check using the specific linkage matrices in Table \ref{links242} that each of the four terms
 in (\ref{b1test}) are equal.  This verifies three of the 42 spin-consistency conditions in this case.
 It is similarly easy to check that we get exactly the
 same matrix for each of the four terms when we compute using the opposite relative sign in each
 of the matrix sums in (\ref{b1test}). (The reader should find this an instructive
 exercise.) This adds four more checks, so that we have thus
 verified
 seven of the 42 spin-consistency conditions.  If we do a similar thing for each of the
 six Lorentz generators specified by (\ref{mz}) we readily verify all $6\times 7=42$ requirements of
 our
 third sieve; direct computation using the matrices in Table \ref{boorots} and Table \ref{links242}
 does show that our spin consistency test is affirmative in each case.  This is an interesting non-trivial
 check corroborating the ability of the linkage matrices to enhance
 to four-dimensions.

 \begin{table}
 \begin{center}
 \begin{tabular}{cc}
 ${\cal B}^1 \,\,=\,\, \fr12\,\ba{cc|cc}&&1&\\&&&1\\\hline 1&&&\\&1&& \ea$ &
 ${\cal R}_1 \,\,=\,\, \fr12\,\ba{cc|cc}&&&-1\\&&1&\\\hline &-1&&\\1&&&\ea$ \\[.5in]
 ${\cal B}^2 \,\,=\,\, \fr12\,\ba{cc|cc}&&&1\\&&-1& \\\hline &-1&& \\1&&& \ea$ &
 ${\cal R}_2 \,\,=\,\, \fr12\,\ba{cc|cc} &&1& \\&&&1\\\hline -1&&& \\&-1&& \ea$ \\[.5in]
 ${\cal B}^3 \,\,=\,\, \fr12\,\ba{cc|cc}1&&&\\&1&&\\\hline &&-1& \\&&&-1 \ea$ &
 ${\cal R}_3 \,\,=\,\, \fr12\,\ba{cc|cc}&-1&&\\1&&&\\\hline &&&1\\&&-1& \ea$
 \end{tabular}
 \caption{Lorentz boost and rotation operators in the spinor
 representation using a Majorana basis.}
 \label{boorots}
 \end{center}
 \end{table}

 We may now use (\ref{b1test}) to determine the component spin representation
 matrices. Direct computation shows that $(\,\tilde{B}_1\,)_{\hat{\imath}}\,^{\hat{\jmath}}$ has the same
 $4\times 4$ matrix form as $(\,{\cal B}_1\,)_A\,^B$ specified in
 Table \ref{links242}.  The index structures differ, however, because
 the fields $\psi_{\hat{\imath}}$ and the ambient supercharges
 $Q_A$ span {\it a priori} distinct vector spaces.
 After we repeat a similar analysis for each of the remaining five equations described
 by (\ref{mz}), we find that all three boost matrices
 $B_a$ and all three rotation matrices $R_a$ are determined to take the
 same $4\times 4$ matrix form as the respective matrices ${\cal B}_a$ and ${\cal R}_a$.
 Since the latter collectively generate a spinor representation, it follows that
 the $B_a$ and $R_a$ generate collectively the same spinor
 representation.
 (This is because they are described by the same specific matrices.)
 In this way we have shown that the vector space spanned by the supercharges $Q_A$ and the
 ostensibly distinct vector space spanned by the
 fermion components $\psi_{\hat{\imath}}$ are isomorphic; \ie, that
 the fermions assemble into a spinor.

 Now, we can generalize our analysis to cases where both the
 bosons and the fermions have potentially non-trivial spin assignments.
 If we allow for non-vanishing $B_a$ and non-vanishing $R_a$, then
 equation (\ref{stan}) generalizes to
  \brr (\,{\cal B}_a\,)_A\,^B\,(\,d_B\pm u_B\,) &=&
      (\,d_A\mp u_A\,)\,\tilde{B}_a
      -B_a\,(\, d_A\mp u_A\,)
      \nonumber\\[.1in]
      (\,{\cal R}_a\,)_A\,^B\,(\,d_B\pm u_B\,) &=&
      (\,d_A\pm u_A\,)\,\tilde{R}_a
      -R_a\,(\,d_A\pm u_A\,)
      \,,
 \label{stan2}\err
 where the component indices have been suppressed.  Note that
 equation (\ref{stan}) is obtained from equation (\ref{stan2}) by setting
 $B_a=0$ and $R_a=0$. This reduces the problem of identifying the
 generating matrices $B_a$, $R_a$, $\tilde{B}_a$, and $\tilde{R}_a$
 to straightforward linear algebra.

 Since the matrices $d_A\pm u_A$ are non-singular, we can use elementary matrix
 algebra to re-organize (\ref{stan2}) into
 an equivalent set of six equations which each involve exactly one of the unknown Lorentz
 generators.  For example, the fermion boost generators are
 determined by the equation
 \brr [\,M_{AB}\,,\,\tilde{B}_a\,]_{\hat{\imath}}\,^{\hat{\jmath}} &=&
      (\,N_{a\,AB}\,)_{\hat{\imath}}\,^{\hat{\jmath}} \,,
 \label{commat}\err
 where $M_{AB}$ and $N_{a\,AB}$ are sets of specific matrices determined by the
 linkage matrices codifying a given one-dimensional supermultiplet, as
 \brr M_{AB}
      &=& (\,d_A-u_A\,)^{-1}\,(d_B+u_B\,)
      \nonumber\\[.1in]
      N_{a\,AB} &=&
      (\,{\cal B}_a\,)_B\,^C\,(\,d_A-u_A\,)^{-1}\,(\,d_C-u_C\,)
      \nonumber\\[.1in]
      & & +(\,{\cal
      B}_a\,)_A\,^C\,(\,d_A-u_A\,)^{-1}\,(\,d_C+u_C\,)\,
      (\,d_A-u_A\,)^{-1}\,(\,d_B+u_B\,) \,,
 \label{matdefs}\err
 where a sum is implied over the repeated $C$-index but not over the repeated $A$-indices.
 Each of the matrices specified in (\ref{matdefs}) are straightforward to
 compute using a given set of linkage matrices and a selected
 four-dimensional spin structure.  Each of the matrix equations
 (\ref{commat}) are then well-posed linear algebra problems.
 For each choice of the $a$-index, (\ref{commat}) supplies a
 separate matrix equation for each choice of the spinor indices $A$
 and $B$.  The requirement that each of these equations provides an
 identical solution for $\tilde{B}_a$ provides the generalized
 statement of the spin-consistency ``third sieve" requirement
 discussed above in the simpler setting involving de-facto scalar bosons.

 The way to solve (\ref{commat}) is determined by elementary linear algebra.
 Each map $X_{\hat{\imath}}\,^{\hat{\jmath}}\to [\,M_{AB}\,,\,X\,]_{\hat{\imath}}\,^{\hat{\jmath}}$ describes a
 linear transformation that can be written as a new matrix, known as the adjoint transformation for
 $(\,M_{AB}\,)_{\hat{\imath}}\,^{\hat{\jmath}}$, called
 $(\,{\rm ad}\,M_{AB}\,)_{\hat{\imath}}\,^{\hat{\jmath}}{}_{\hat{k}}\,^{\hat{l}}$. The solutions to (\ref{commat}) are obtained by solving the
 new matrix equation
 \brr (\,{\rm ad}\,M_{AB}\,)_{\hat{\imath}}\,^{\hat{j}}{}_{\hat{k}}\,^{\hat{l}}\,(\,\tilde{B}_a\,)_{\hat{l}}\,^{\hat{k}}
      &=&  (\,N_{a\,AB}\,)_{\hat{\imath}}\,^{\hat{\jmath}}
 \label{adeq}\err
 by row reduction. This provides an algorithm readily implementable by computer.
 Part of the question of whether a compatible Lorentz
 structure exists becomes the question of whether (\ref{adeq}) has a
 solution, providing a ``fourth sieve".\footnote{Some of our
 sieve-like consistency requirements described in
 \cite{Enhancement} and in this paper may be redundant.  We
 leave this as future work to sort out a minimalist statement of
 spin-consistency.}

 A procedure similar to the one we have outlined for determining
 $\tilde{B}_a$ may be brought to bear on each of the remaining
 Lorentz generators, $\tilde{R}_a$, $B_a$, and $R_a$.  The relevant
 matrix equation analogous to (\ref{commat}) is obtained by
 straightforward algebraic manipulation of (\ref{stan2}).
 In each case, this provides a multiplicity of linear algebra
 problems analogous to those codified by the different possible choices of the spinor indices $A$
 and $B$ in (\ref{commat}).  By demanding consistency among the solutions
 for each case we generalize the spin-consistency problem described
 above.  For consistent cases, the relevant spin generators are
 determined by the same technique as described above in the case of
 $\tilde{B}_a$.

 There is another layer of subtlety to our algorithm.  In fact, equations such as (\ref{adeq})
 typically have more than one solution.  So, in principle, we need to
 keep track of all solutions for each of the six boson spin generators
 $(\,B_a\,,\,R_a\,)$ and each of the six fermion spin generators
 $(\,\tilde{B}_a\,,\,\tilde{R}_a\,)$.  Then we need to enforce that
 the spin algebra is satisfied on each set.  In the general case, this
 would require systematic checking of the various possible
 combinations.  We are developing software to attend to this task,
 and are eager to see how this plays out in specific examples.  We
 expect to have more to say about this in the future.

 In conclusion, we have explained a practicable algorithm for resolving the full
 implications implied by the imposing Lorentz invariance
 on postulate enhancements of one-dimensional linkage matrices.
 This provides a specific way to
 test whether spin structures can be consistently assigned to
 enhancements of one-dimensional supermultiplets. In cases where our
 spin-consistency conditions are satisfied, the method described above
 allows one to actually compute spin representation assignments of the enhanced
 component fields.  We have shown also how this technique simplifies
 into tidy closed-form expressions for the component fermion spin
 generators in cases where the bosons are postulated to be scalars.
 This presentation adds an interesting ingredient to the paradigm
 described in \cite{Enhancement}, and explains how information about
 ambient spin assignments are accessibly coded in one-dimensional
 shadows.

 \vspace{.1in}

 \noindent{\bf Acknowledgements}\\[.1in]
 The authors are grateful to Kevin Iga for beneficial discussions,
 and also to Charles Doran, Tristan H{\"u}bsch, and S.~J.~Gates, Jr.
 for collaborative work which precipitated this paper.
 M.F. is thankful to the Slovak Institute for Basic Research (SIBR), in Podva${\check{\rm z}}$ie
 Slovakia , where much of this
 work was completed, for providing hospitality, peace, love, and halu$\check{\rm s}$ky.

 \Refs{References}{[00]}
 \bibitem{Enhancement}
 M.~G.~Faux, K.~M.~Iga, and G.~D.~Landweber:
 {\em Dimensional Enhancement via Supersymmetry},
 arXiv:hep-th/0907.3605;
 \Bib{FG1} M.~Faux and S.~J.~Gates, Jr.:
 {\em Adinkras: A Graphical Technology for Supersymmetric Representation Theory},
  Phys.~Rev.~{\bf D71}~(2005),~065002;
   \Bib{T2}
   Z.~Kuznetsova, M.~Rojas, F.~Toppan,
   {\em Classification of irreps and invariants of the $N$-extended
   supersymmetric quantum mechanics},
   JHEP 03 (2006) 098;
 \bibitem{GatesRana}
 S.J. Gates, Jr. and L. Rana:
 {\em A Theory of Spinning Particles for Large N Extended
 Supersymmetry},
 Phys.~Lett.{\bf B352} (1995) 50-58,
 hep-th/9504025\,;
 S.J. Gates, Jr. and L. Rana:
 ``A Theory of Spinning Particles for Large N Extended
 Supersymmetry (II)",
 Phys.~Lett. {\bf B369} (1996) 262-268,
 hep-th/9510151;
 \bibitem{Enuf}
 S. J. Gates, Jr., W.D. Linch, III and
 J. Phillips:
 {\em When Superspace is Not Enough},
 hep-th/0211034;
 \bibitem{Filtered}
 C.~Doran, M.~Faux, S.~J.~Gates, Jr., T.~H{\"u}bsch, K.~Iga,
 G.~Landweber:
 {\em Off-Shell supersymmetry and filtered Clifford
 supermodules}, arXiv:math-ph/0603012v2 (in progress);
  \Bib{DFGHIL01} C.~Doran, M.~Faux, S.~J.~Gates, Jr., T.~H{\"u}bsch, K.~Iga, G.~Landweber:
 {\em On Graph Theoretic Identifications of
 Adinkras, Supersymmetry Representations and Superfields},
  Int. J. Mod. Phys. {\bf A}22 (2007) 869-930;
 \Bib{RETM}
 C.~Doran, M.~Faux, S.~J.~Gates, Jr., T.~H{\"u}bsch, K.~Iga, G.~Landweber:,
 {\it On the matter of $N=2$ matter},
 Phys. Lett. {\bf B}659 (2008) 441-446.
 \endRefs

 \end{document}